\documentclass[11pt]{article}
\usepackage[a4paper,total={6.8in, 10.1in}]{geometry}
\usepackage{authblk}
\usepackage{longtable}
\usepackage{booktabs}
\usepackage{amsmath}
\usepackage{hyperref}
\usepackage{graphicx}
\usepackage{xcolor}
\usepackage[backend=biber,style=numeric-comp,sorting=none]{biblatex}
\usepackage[font=small,labelfont=bf]{caption}

\usepackage{float}
\usepackage{tabularx}
\usepackage{multirow}
\usepackage{array}
\usepackage{placeins}
\usepackage{caption}
\usepackage{xurl}
\usepackage{graphicx}
\usepackage{lineno}
\usepackage{hyperref}
\usepackage{makecell}
\title{\textbf{Enhancer–promoter proximity predicts transcriptional competence but not transcriptional output in the \textit{Drosophila} brain}}
\author[1,4]{Olivier Messina}
\author[2,5]{Loucif Remini}
\author[3]{Christopher H. Bohrer}
\author[1]{Jean-Bernard Fiche}
\author[2]{Jean-Charles Walter}
\author[2]{Andrea Parmeggiani}
\author[1,*]{Marcelo Nollmann}
\affil[1]{Centre de Biologie Structurale, University of Montpellier, INSERM, CNRS, Montpellier, France}
\affil[2]{Laboratoire Charles Coulomb (L2C), University of Montpellier, CNRS, Montpellier, France}
\affil[3]{Laboratory of Receptor Biology and Gene Expression, Center for Cancer Research, National Cancer Institute, National Institutes of Health, United States}
\affil[4]{Currently at Center for Integrative Genomics, University of Lausanne, 1015 Lausanne, Switzerland.}
\affil[5]{Currently at University of Geneva, Quai Ernest Ansermet 30, 1205 Geneva, Switzerland} 

\affil[*]{Corresponding author: marcelo.nollmann-martinez@cnrs.fr}

\date{}

\begin{document}

\maketitle

\begin{abstract}
How 3D genome architecture contributes to transcriptional specificity across neuronal cell types remains unclear. Here, we used multiplexed chromatin tracing to map chromatin architecture and cell identity at single-cell resolution in the adult \textit{Drosophila} brain. We found that enhancer–promoter (E--P) proximity was increased in transcriptionally active compared with inactive neurons. Analysis of single traces revealed the existence of distinct proximal and distal E--P states, with active neurons enriched in the proximal state. However, this relationship broke down across active neuronal subtypes, where neither E--P proximity nor chromatin accessibility predicted transcriptional output. Thus, 3D genome organization distinguishes transcriptionally competent from inactive neuronal states without quantitatively specifying transcriptional output. Our findings support a model in which E--P proximity establishes a permissive structural state, while additional cell-type-specific regulatory mechanisms tune transcriptional output.
\end{abstract}

\section*{Introduction}

\noindent The brain consists of a myriad of specialized neuronal and glial cell types, each characterized by distinct morphology and function, which collectively enable complex behaviors and cognitive processes such as learning and memory. The \textit{Drosophila melanogaster} brain, with its relatively compact size of approximately 100,000 cells \cite{Kremer2017}, offers an exceptional model for exploring the mechanisms underlying neuronal diversity and function. Recent advances in whole-brain annotation and connectomics have revealed the intricate morphology, connectivity, and function of hundreds of neuronal types \cite{Schlegel2024}. Complementing these efforts, single-cell omics technologies have enabled detailed mapping of transcriptomic and chromatin accessibility profiles across neuronal populations in the \textit{Drosophila} brain \cite{Davie2018,Janssens2022}. These studies have shed light on the gene regulatory networks that establish and maintain neuronal identity. However, how these regulatory programs are encoded in the three-dimensional (3D) organization of the genome in individual neuronal cell types remains largely unexplored. In particular, how distal enhancers spatially communicate with their target genes to generate cell-type-specific transcriptional programs in the adult brain remains poorly understood.\\

\noindent An essential component of distal gene regulation is the 3D organization of the genome. Eukaryotic chromosomes are partitioned into topologically associating domains (TADs) \cite{Dixon2012-ti,Nora2012-yj,Sexton2012-jv}, which frequently encompass gene promoters (P) and their enhancers (E) and contribute to defining regulatory landscapes \cite{Schwarzer2014-lk,De_Laat2013-qf}. Within TADs, specific enhancer--promoter (E--P) configurations provide an additional level of spatial organization associated with gene regulation. Notably, substantial chromatin reorganization occurs during neural differentiation in both mammals \cite{Bonev2017,Titus2023,Monahan2019-sw} and flies \cite{Mohana2023,Pollex2024}. For instance, new regulatory interactions, including neuron-specific E--P contacts, emerge in post-mitotic neurons compared with progenitor cells. These observations highlight the importance of 3D genome architecture in establishing and maintaining neuronal identity and function.\\

\noindent Despite the importance of E--P communication for transcriptional regulation, studies have revealed diverse and sometimes contradictory relationships between E--P proximity and transcription depending on developmental and transcriptional context. In some cases, high frequencies of E--P spatial co-localization are restricted to developmental stages when target genes are actively transcribed, consistent with an instructive mode of enhancer action \cite{Yang2024-jx,Chen2024-ck}. During early development, permissive E--P loops can exist independently of transcriptional activity \cite{Espinola2021-mw,Ing-Simmons2021}, facilitating rapid gene activation \cite{Misteli2021-pl}. At later stages of embryonic development, E--P proximity can act either permissively or instructively \cite{Pollex2024}, while E--P distance can also increase during transcriptional activation associated with cell differentiation \cite{Benabdallah2019,Alexander2019}. These observations raise two related but distinct questions: whether E--P proximity distinguishes transcriptionally competent from inactive neuronal states, and whether differences in proximity quantitatively determine transcriptional output once a gene is active. The latter distinction is particularly relevant in the brain, where related neuronal subtypes can share transcriptional programs while expressing individual genes at different levels. Addressing these questions requires direct measurements of 3D chromatin organization across defined neuronal populations with distinct transcriptional states, while retaining single-cell resolution and cellular identity within the adult brain.\\

\noindent Here, we address these questions using the \textit{rutabaga} (\textit{rut}) locus as a model of E--P regulation in the post-mitotic brain. \textit{rut} is expressed at different levels across defined neuronal populations, providing a system to investigate both the transition between inactive and active transcriptional states and quantitative differences in transcriptional output among active neurons. We used Hi-M, an imaging approach combining chromatin tracing (multiplexed DNA-FISH) with cell-type identification in cryosectioned adult \textit{Drosophila} brain tissue, to reconstruct 3D genome organization in defined neuronal populations at single-cell resolution. This approach resolves chromatin organization from TAD-scale folding to individual E--P configurations. We find that E--P proximity distinguishes transcriptionally active from inactive neurons and reflects a redistribution between distinct promoter-proximal and distal states. However, among active neuronal subtypes, neither E--P proximity nor chromatin accessibility quantitatively predicts transcriptional output. These findings support a model in which 3D chromatin organization establishes transcriptional competence, while transcriptional output is quantitatively tuned by additional regulatory mechanisms.

\section*{Results}

\subsection*{High-resolution Hi-M enables visualization of 3D chromatin structure in the adult fly brain}

\begin{figure}[H]
  \centering
  \includegraphics[height=0.78\textheight]{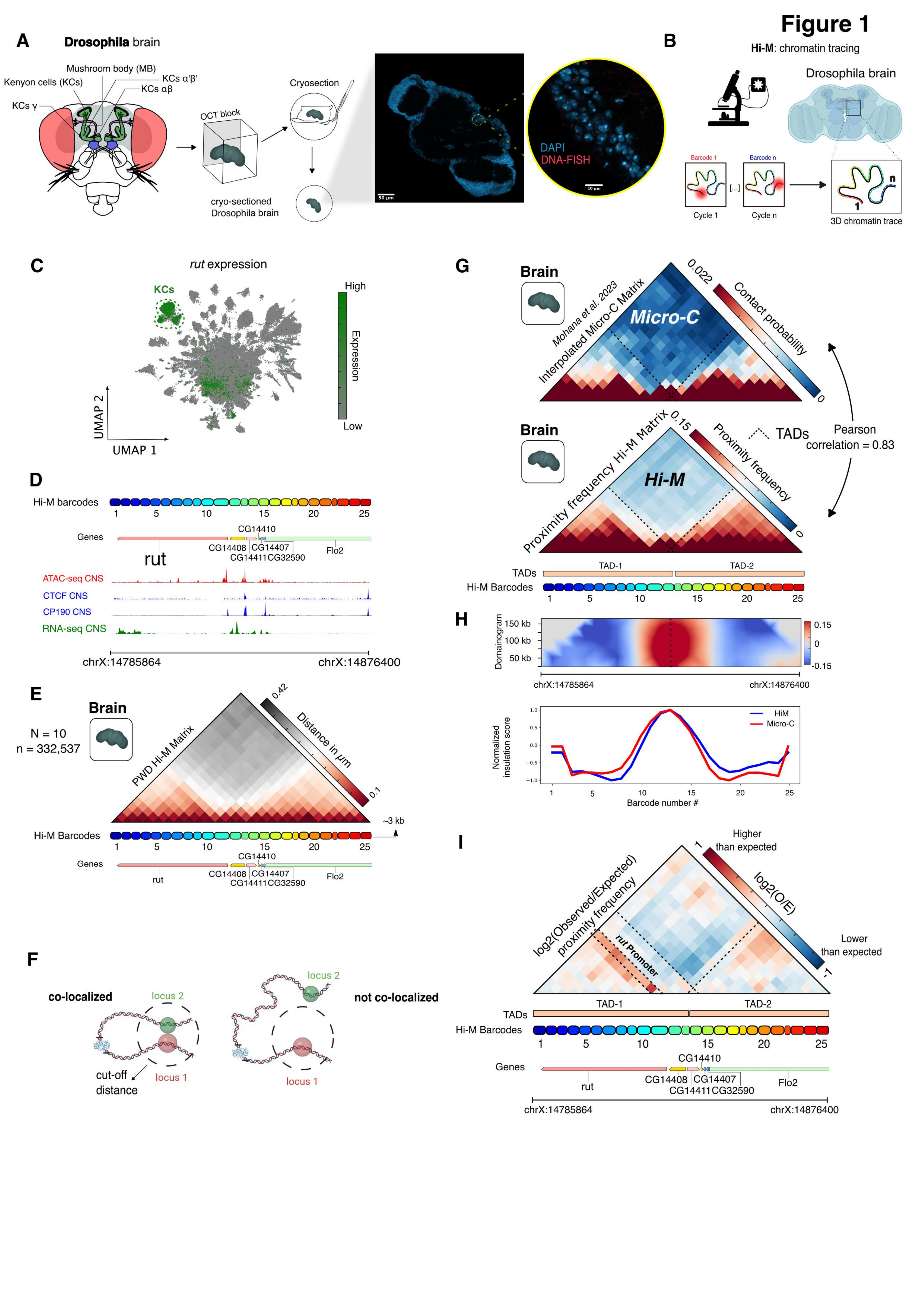}
  \captionsetup{
    width=\textwidth,
    justification=justified,
    singlelinecheck=false
  }
  \label{fig:fig1}
  \vspace{-80pt}
  \caption{%
 \textbf{Hi-M enables visualization of chromosome 3D structure in single cells in cryosectioned \textit{Drosophila} brains.}
    \textbf{A} Schematic of the \textit{Drosophila} head highlighting the mushroom body (MB) and its Kenyon cell (KC) subtypes: $\alpha\beta$, $\alpha'\beta'$, and $\gamma$. The brain was fixed and embedded in OCT prior to cryosectioning.\textbf{B} Schematic of the imaging-based strategy used to trace chromatin architecture at the single-cell level in cryosectioned \textit{Drosophila} brains (Hi-M). \textbf{C} UMAP visualization of adult \textit{Drosophila} brain cell-types based on snRNA-seq data. \textit{rut} expression is shown as a green gradient, and the KCs cluster is highlighted. \textbf{D} Top: Barcodes used for Hi-M sequential imaging are shown as color-coded boxes along the \textit{rut} locus (chrX:14785864-14876400, dm6). Middle: Gene locations.  Bottom: Tracks showing chromatin accessibility (ATAC-seq), insulator protein binding (CTCF and CP190 ChIP-seq), and gene expression (RNA-seq) in the CNS, displayed with genomic coordinates. \textbf{E} Hi-M median PWD matrix of the \textit{rut} locus from cryosectioned \textit{Drosophila} brain tissue, constructed from n = 332,537 traces (N = 10 experiments). Schematic of chromatin interactions between two loci defined by a distance  cut-off. Loci separated by less than the  cut-off radius are considered co-localizing. \textbf{F} Schematic illustrating the use of a  cut-off distance to define co-localization between Hi-M barcodes.\textbf{G} Top : Micro-C contact matrix in \textit{Drosophila} CNS along the \textit{rut} locus. Bottom : Proximity frequency Hi-M matrix generated with a  cut-off distance of 150 nm. TADs identified in Hi-C data are shown as dashed lines. \textbf{H} Top: Domainogram showing insulation scores derived from Hi-M data at different window sizes (see \hyperref[sec:methods]{Methods}). Bottom: Comparison of normalized insulation scores derived from Hi-M (blue) and Micro-C (red) data along the \textit{rut} locus. \textbf{I} Hi-M distance-normalized proximity frequency matrix. Red indicates higher-than-expected proximity and blue indicates lower-than-expected proximity. The \textit{rut} promoter bin is shown as a dashed rectangle.\\
  }
\end{figure}

\noindent To examine whether 3D chromatin organization changes across post-mitotic neuronal subtypes in the adult \textit{Drosophila} brain, we performed Hi-M imaging in brain cryosections at ~3kb resolution (Figs. \hyperref[fig:fig1]{1A-1B}) \cite{Bardou2026-iw}. We focused specifically on the \textit{rutabaga} (\textit{rut}) gene, which encodes a Ca2+/Calmodulin-responsive adenylyl cyclase involved in learning and memory \cite{Levin1992}. \textit{rut} is specifically expressed in the three Kenyon Cell (KCs) types of the mushroom body (MB), a brain neuropil involved in olfactory learning and memory \cite{McGuire2001,Heisenberg2003,Aso2014} (Figs. \hyperref[fig:fig1]{1C} and \hyperref[fig:fig1]{S1A}). The Hi-M oligopaints library contained 25 genomic segments (barcodes) spanning a 90 kb region containing the \textit{rut} promoter (Fig. \hyperref[fig:fig1]{1D}) (\hyperref[sec:methods]{Methods}).\\

\noindent The Hi-M dataset contained over 4.5 million distance pairs, across 332,537 single traces from 10 independent replicates \cite{Bardou2026-iw}. The reconstructed median pairwise distance matrices (PWD) from the 10 replicates were highly correlated and reproducible (Fig. \hyperref[fig:figS1]{S1B}). Typical Hi-M benchmarks, such as barcode detection efficiencies, bootstrapping analysis, and distance histograms for all barcode pairs, are provided in Figs. \hyperref[fig:figS1]{S1C-D}) and in Ref. \cite{Bardou2026-iw}. To correct tracing errors, we applied Loci Enabled Advanced Resolution (LEAR) \cite{Bohrer2025-nz}, which leverages localization-error-dependent inflation of pairwise displacement variances to iteratively approximate accurate loci positions (see \hyperref[sec:methods]{Methods}). By calculating the median PWD for each pair of genomic regions, we obtained an ensemble view of chromatin organization at this locus (Fig. \hyperref[fig:fig1]{1E}). As expected, regions located at closer genomic distances showed lower PWD distances (Fig. \hyperref[fig:fig1]{1E}), with a dependence between physical and genomic distance that followed a power law (Fig. \hyperref[fig:figS2]{S2A}).\\

\noindent Next, we compared these results to existing Micro-C data from the entire \textit{Drosophila} central nervous system (CNS) \cite{Mohana2023}. For this, we computed the proximity frequency for any two given chromatin regions by calculating the number of times two loci co-localized within a given cut-off distance (Fig. \hyperref[fig:fig1]{1F}). By calculating this metric for all combinations of chromatin regions, we constructed frequency proximity maps for different cut-off distances and compared each map to the publicly available Micro-C CNS dataset using Pearson correlation (Fig. \hyperref[fig:figS2]{S2B}) (see \hyperref[sec:methods]{Methods}). The highest Pearson correlation ($\Gamma=0.83$) was found at a cut-off distance of 150 nm, which corresponds to cut-off distance values typically used in the literature (\cite{Cardozo_Gizzi2019-lc,Mateo2019-wg}). As expected, using this optimal cut-off distance, we observed a very high similarity between the Micro-C map and the ensemble Hi-M proximity frequency matrix (Fig. \hyperref[fig:fig1]{Fig.1G}).\\

\noindent The Hi-M map of this locus displays two genomic regions with higher proximity frequencies (dashed lines in Fig. \hyperref[fig:fig1]{1G}), as quantitatively confirmed by domainogram and insulation score analysis (Fig. \hyperref[fig:fig1]{1H}, \hyperref[sec:methods]{Methods}). These two regions correspond to the TADs annotated from Micro-C data (TAD-1 and TAD-2, Fig. \hyperref[fig:fig1]{1G}), and are insulated by multiple CTCF and CP190 binding sites, as well as open chromatin regions, and highly transcribed genes, a common trait of TAD borders in \textit{Drosophila} \cite{Ulianov2016} (Fig. \hyperref[fig:fig1]{1D}).\\

\noindent The transcriptional unit of the \textit{rut} gene is entirely contained within TAD-1, and exhibits preferential intragenic interactions that bring the \textit{rut} promoter more often in closer proximity to its gene body elements. To test this, we normalized the proximity map using the proximity frequency versus genomic distance curve to determine whether loci co-localized more often than expected (Fig. \hyperref[fig:figS2]{S2C}) (\hyperref[sec:methods]{Methods}). Notably, we observed a significant enrichment of proximity frequencies between the \textit{rut} promoter and its gene body, even above that caused by TADs themselves (Fig. \hyperref[fig:fig1]{1I}). In summary, the implementation of chromatin tracing in the adult fly brain detects preferential interactions between the rut promoter and regulatory elements located within its gene body, despite a relatively low proportion of cells expressing \textit{rut} in the fly brain ($\approx5\%$ of neurons) \cite{Cognigni2018-vw, Meinertzhagen2018-lh}.\\

\subsection*{E--P proximity distinguishes transcriptionally active and inactive neuronal populations}

\begin{figure}[H]
  \centering
  \includegraphics[height=0.85\textheight]{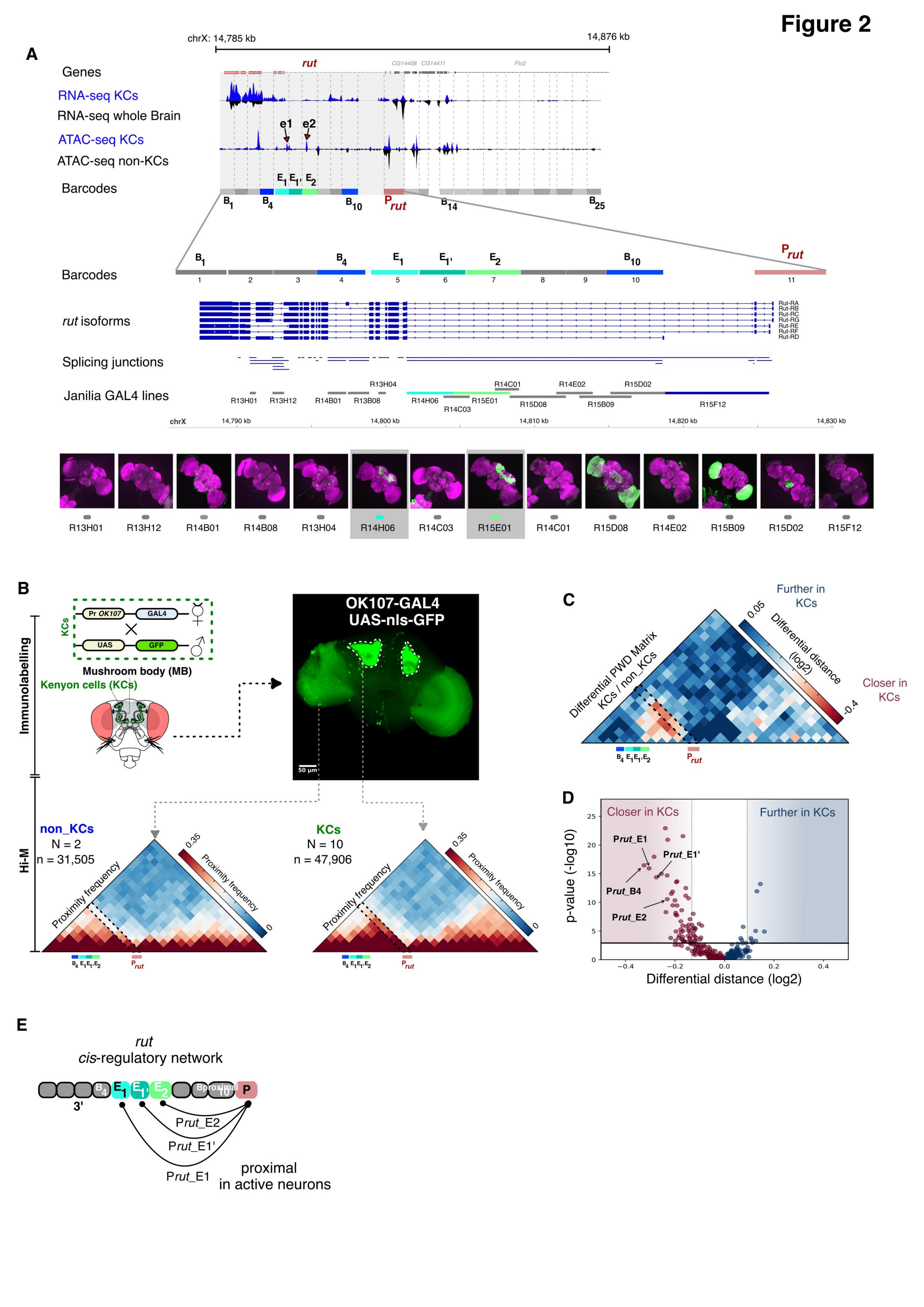}
  \captionsetup{
    width=\textwidth,
    justification=justified,
    singlelinecheck=false
  }
  \vspace{-40pt}
  \caption{%
    \textbf{Cell-type-specific E--P interactions correlate with differential \textit{rut} expression in Kenyon cells}.
   \textbf{A} Tracks showing gene expression levels (RNA-seq) and differential chromatin accessibility (ATAC-seq) in Kenyon cells (KCs, blue) versus the rest of the brain (black). Zoom-in of the Hi-M library between barcode 1 (B1) and barcode 11 (Prut) is shown, along with the different \textit{rut} isoforms. Fourteen screened CREs are displayed in the zoomed region, together with GFP expression for these CREs along the \textit{rut}, as conducted by the Janelia FlyLight Project Team (see \hyperref[sec:methods]{Methods}). \textbf{B} Schematic illustrating the immuno-HiM strategy. Left: Schematic of the UAS-GAL4 system used to drive GFP expression in the mushroom bodies. Right: Confocal image showing the GFP signal in the \textit{Drosophila} adult brain. Bottom: Hi-M proximity frequency matrix along the \textit{rut} locus in non-KCs (left) and KCs (right). \textbf{C} Differential PWD matrix comparing KCs and non-KCs cells. \textbf{D} Volcano plot showing the log2(KCs/ non-KCs) PWD distance (x-axis) and -log10 (P-value). E--P interactions are highlighted with black arrows. \textbf{E} Schematic representation of \textit{rut}'s \textit{cis}-regulatory network and regions proximal in active neurons. 
  }
  \label{fig:fig2}
\end{figure}

To shed light on these preferential proximity patterns, we analyzed how transcription and chromatin accessibility changed between KCs and non-KCs at the \textit{rut} locus. For this, we generated pseudo-bulk transcription and accessibility profiles using existing snRNA-seq and snATAC-seq datasets \cite{Janssens2022} by aggregating the profiles of KC-$\alpha\beta$, KC-$\alpha'\beta'$ and KC-$\gamma$ into KC-specific transcription and chromatin accessibility, while similarly combining non-Kenyon cells (non-KC) to reconstruct the transcriptional and accessibility profiles of the remaining CNS cell-types (Fig. \hyperref[fig:fig2]{2A}). As expected, \textit{rut} is highly expressed in KCs, and displays very low levels of transcription in non-KCs, whereas other genes in the locus show the opposite trend (e.g. \textit{CG14408}, Fig. \hyperref[fig:fig2]{2A}, RNA-seq profiles). We also observed multiple differentially accessible regions (DARs) located within TAD-1 and specifically accessible in KCs. From these, only two DARs ($e_1$, $e_2$, Fig. \hyperref[fig:fig2]{2A}, red arrows) fall within non-coding regions, and drive reporter gene expression in the MB (R14H06, R15E01, respectively; Fig. \hyperref[fig:fig2]{2A}, bottom panel). Thus, despite the presence of multiple DARs at the \textit{rut} locus, only $e_1$ and $e_2$ (contained within barcodes E1, E1' and E2) display specific enhancer activity in mushroom body KCs.\\

\noindent To investigate whether these enhancers were closer to the \textit{rut} promoter specifically in KCs, we analyzed the relative E--P 3D proximities in KCs versus non-KCs. For this, we combined Hi-M with green fluorescent protein (GFP) immunostaining-driven detection of specific fly brain cell-types using distinct GAL4/UAS-nls-GFP lines \cite{Bardou2026-iw}. Using the OK107-GAL4 transgene, we drove nls-GFP expression in the MB, allowing us to distinguish KCs from non-KCs based on nls-GFP signal (Fig. \hyperref[fig:fig2]{2B}, top panels, \hyperref[sec:methods]{Methods}). Using this signal, we classified chromatin traces into KCs and non-KCs, and used them to reconstruct ensemble proximity frequency maps for these two cell-types. The overall chromatin structure of the locus was similar between KCs and non-KCs; however, we observed notable differences in intra-TAD proximities, particularly between the \textit{rut} promoter and other regions within TAD-1 (Fig. \hyperref[fig:fig2]{2B}, dashed lines).\\

\noindent To better visualize these differences, we plotted the differential median PWD map between KCs and non-KCs by calculating the $\log_2(PWD_{KCs}/PWD_{non-KCs})$ (Fig. \hyperref[fig:fig2]{2C}). In this representation, regions closer in KCs appear in red, while regions further in KCs appear in blue. In KCs, the \textit{rut} promoter is closer to the barcodes containing the identified KC enhancers $e_1$ and $e_2$ ($E_1$, $E_1'$, $E_2$). We evaluated the statistical significance of these differences by leveraging the single-cell nature of Hi-M. Specifically, for each bin in the proximity map we plotted the differential distance $\log_2(PWD_{KCs}/PWD_{non-KCs})$ as a function of a p-value calculated from single-cell PWD variations (\hyperref[sec:methods]{Methods}). This analysis further demonstrates that the distances between the \textit{rut} promoter and its enhancers, specifically $P_\textit{rut}-E_1$, $P_\textit{rut}-E_{1'}$, and $P_\textit{rut}-E_2$, are significantly lower in \textit{rut}-expressing cells (KCs) (Fig. \hyperref[fig:fig2]{2D}) than in non-expressing cells (non-KCs). \\

\noindent Direct measurements of E--P proximity frequencies in KCs and non-KCs (Fig. \hyperref[fig:figS3]{S3A}) indicate that \textit{rut}'s putative enhancers are not in permanent contact to the promoter ($\approx25-30\%$), even in cells displaying specific \textit{rut} transcription. Notably, we also observe other significant changes in proximity, bringing the promoter closer to other regions of \textit{rut}'s transcriptional unit containing alternative promoters and exon-intron junctions but no enhancer activity (e.g. B4 or B10, Fig. \hyperref[fig:fig2]{2C}, dashed lines, and Fig. \hyperref[fig:figS3]{S3A}). Overall, these experiments and analyses are consistent with an instructive mode of action, where enhancers get in close physical proximity with the promoter to activate it (Fig. \hyperref[fig:fig2]{2E}). 

\subsection*{A two-state model reveals transcription-dependent enrichment of E--P proximal states}

\begin{figure}[H]
  \centering
  \includegraphics[height=0.85\textheight]{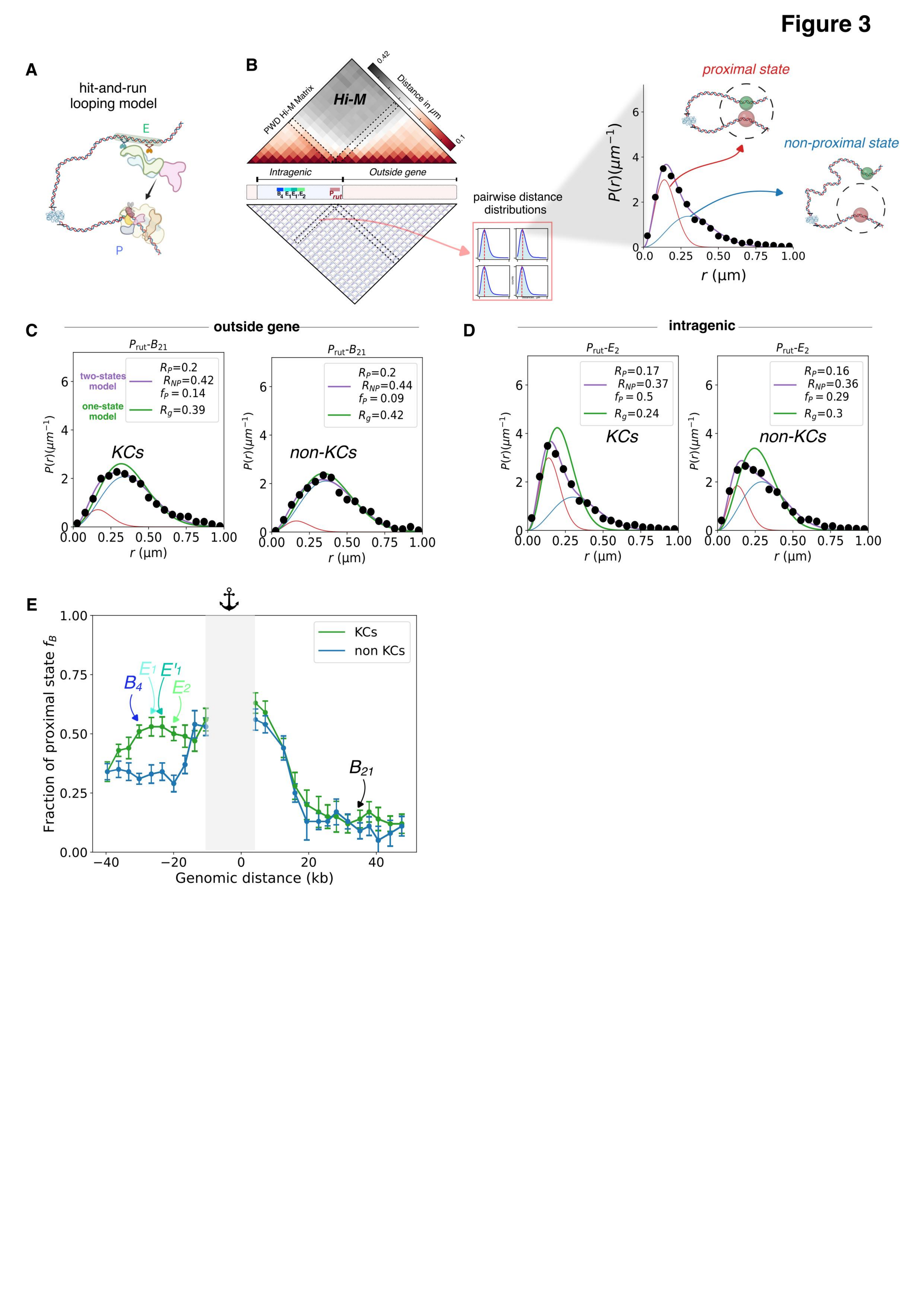}
  \captionsetup{
    width=\textwidth,
    justification=justified,
    singlelinecheck=false
  }
  \vspace{-200pt}
  \caption{%
    \textbf{Two-state enhancer–promoter distance distributions reveal intragenic transcription-dependent states.}.
        \textbf{A} Schematic representation of the hit-and-run model. \textbf{B} Top: Hi-M median PWD matrix of the \textit{rut} locus in adult \textit{Drosophila} brain tissue. Bottom : Histogram matrix of PWD for all barcode pairs. A zoomed-in view of four histograms is shown on the right. The \textit{rut} promoter bin is indicated by a dashed rectangle. "Intragenic" refers to interactions between bins located within the core gene, whereas "outside gene" refers to interactions involving bins located outside the core gene region. \textbf{C, D} Physical distance distributions $P(r)$ between $P_{rut}$ and the $B_{21}$ element outside the gene (\textbf{C}) or the intragenic enhancer $E_2$ (\textbf{D}) in KCs and non-KCs, fitted with one- or two-states models. The red and blue curves in (D) respectively represent the contributions of the proximal and non-proximal states. \textbf{E} Fraction of the proximal state ($f_P$) as a function of genomic distance from the $P_{rut}$ promoter. Profiles compare non-KCs and KCs. Arrows indicate the positions of upstream elements $B_4$, $E_1$, $E'_1$, $E_2$ and the $B_{21}$ element outside the gene.
  }
  \label{fig:fig3}
\end{figure}

\noindent In an instructive model, enhancer activation is expected to involve a promoter-proximal state. If this state is sufficiently long-lived relative to the intrinsic conformational dynamics of the chromatin polymer, E--P distance distributions should be better described by two populations, corresponding to proximal and non-proximal conformational states. Detecting and quantifying these populations would provide a direct estimate of the fraction of alleles in the promoter-proximal state. However, proximity frequencies calculated using an arbitrary distance threshold (Fig. \hyperref[fig:figS3]{~S3A}) depend strongly on the choice of threshold and do not directly distinguish between underlying conformational states \cite{Lleres2025-bf,Yang2024-jx}. To test whether distinct proximal and non-proximal states could be detected in our data, we fitted E--P distance distributions using a two-state model (Fig. \hyperref[fig:fig3]{3B}), as in our previous work \cite{Remini2024-aa,Remini2025-jcp}.\\

\noindent We first focused on distance distributions between $P_{rut}$ and barcodes located outside the gene body (from barcodes B15 to B25). These distributions could be fitted equally well by a single- and by a two-state model, for both KC and non-KCs (Figs. \hyperref[fig:fig3]{3C}, \hyperref[fig:figS3]{S3B}, and \hyperref[fig:figS3]{S3C}); thus, a single state is sufficient to explain the PWD distributions for both KC and non-KC neurons (Fig. \hyperref[fig:fig3]{3C}).\\

\noindent Next, we analyzed PWD distributions between $P_{rut}$ and intragenic regions (barcodes B1--B10). In contrast to promoter-distal pairs, for which a single-state model was sufficient, these distributions could no longer be satisfactorily described by a single-state model (Fig. \hyperref[fig:figS3]{S3B}) and were consistently better fitted by a two-state model in both KCs and non-KCs (Figs. \hyperref[fig:fig3]{3D}, \hyperref[fig:figS3]{S3D}). In these fits, the proximal and non-proximal states are characterized by their relative abundance ($f_P$ and $1-f_P$) and by their characteristic pairwise distances ($R_P$ and $R_{NP}$). \\

\noindent The mean distances associated with each state were similar in KCs and non-KCs, indicating that the proximal and non-proximal states correspond to comparable spatial configurations in transcribing and non-transcribing cells. However, their relative abundances differed markedly. For pairs involving the promoter and a region outside the gene, the fraction of traces in proximal/non-proximal states decreased rapidly with genomic distance and displayed minor changes between KCs and non-KCs (Fig. \hyperref[fig:fig3]{3E}); thus, transcription did not significantly shift the balance between states when the promoter interacts with regions outside the gene.\\

\noindent In stark contrast, promoter--intragenic pairs displayed a strong transcription-dependent enrichment of the proximal state. This effect was particularly pronounced for enhancer-containing regions, where $\approx 50\%$ of alleles occupied the proximal state in KCs. Thus, transcriptional activation is associated not with a change in the geometry of enhancer--promoter contacts, but with a substantial increase in the probability of occupying a promoter-proximal enhancer configuration. These results provide direct evidence for the existence of discrete E--P proximal states at the \textit{rut} locus and demonstrate that they are strongly enriched in transcriptionally active cells. Yet, whether this enrichment merely distinguishes active from inactive cells, or also quantitatively predicts transcriptional output among active neuronal subtypes, remains unclear.

\subsection*{Quantitative transcriptional output is uncoupled from E--P proximity among active neuronal subtypes}

\begin{figure}[H]
  \centering
  \includegraphics[height=0.8\textheight]{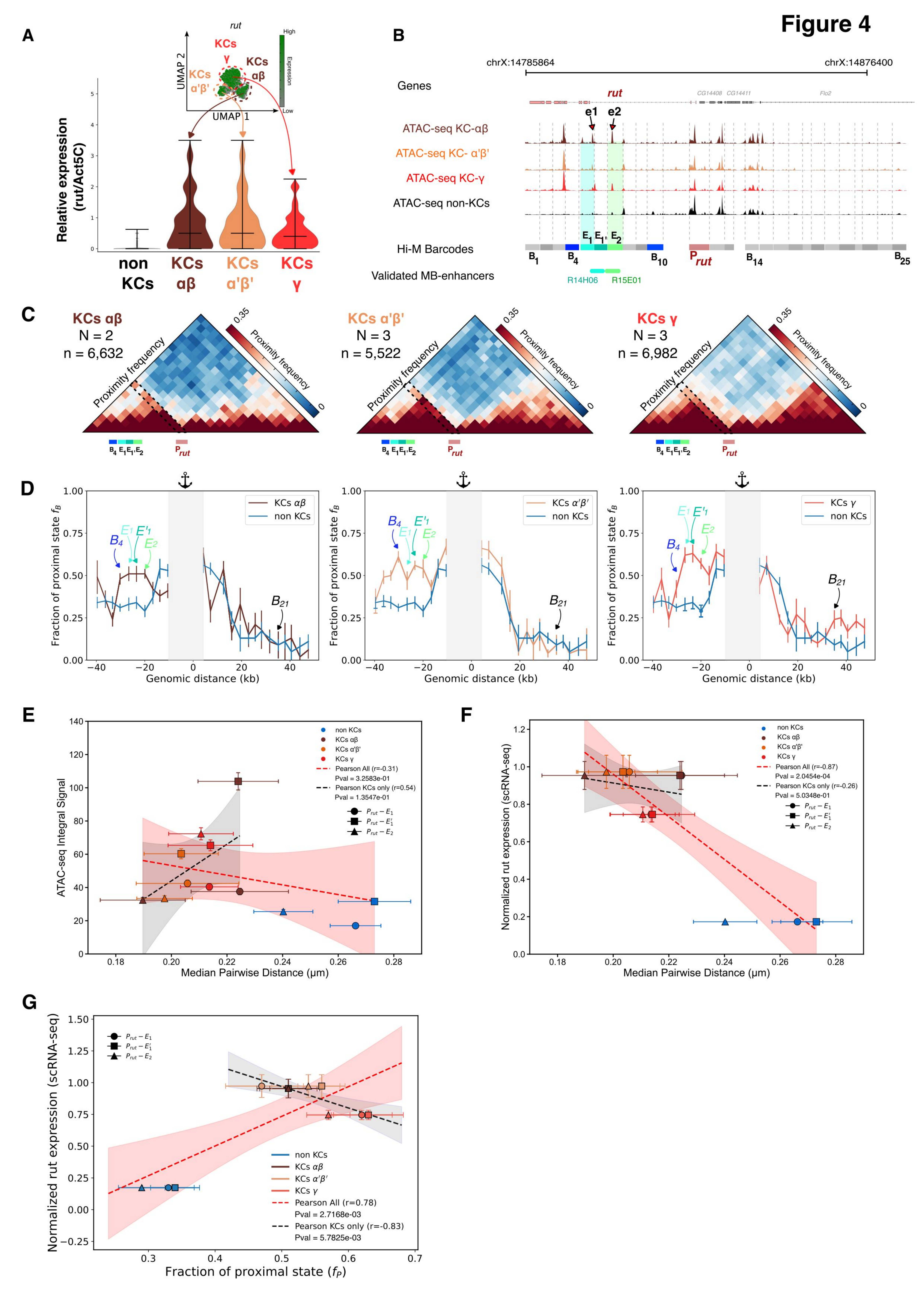}
  \captionsetup{
    width=\textwidth,
    justification=justified,
    singlelinecheck=false
  }
  \vspace{-10pt}
  \caption{
    \textbf{Enhancer–promoter proximity and accessibility are uncoupled from transcriptional output across Kenyon cell subtypes}.
        \textbf{A} Relative \textit{rut} expression levels derived from snRNA-seq experiments for each Kenyon cell subtypes. 
        \textbf{B} Tracks illustrating the chromatin accessibility (ATAC-seq) in Kenyon cell subtypes. 
        \textbf{C} Hi-M proximity frequency matrices for the 3 Kenyon cell subtypes. 
        \textbf{D} Fraction of the proximal state proximal state ($f_P$) as a function of genomic distance from the $P_{rut}$ promoter. Profiles compare non-KCs to three Kenyon cell subtypes: KCs $\alpha\beta$ (left), KCs $\alpha'\beta'$ (middle), and KCs $\gamma$ (right). Arrows indicate the positions of upstream elements $B_4$, $E_1$, $E'_1$, and $E_2$. Error bars indicate the fit uncertainty on $f_P$. 
        \textbf{E} Scatter plot showing the integrated pseudo-bulk ATACseq signal of putative enhancer barcodes across KC subtypes as a function of their median PWD to the \textit{rut} promoter.
        \textbf{F} Scatter plot of median E--P pairwise distances versus \textit{rut}'s mean transcriptional level from pseudo-bulk scRNA-seq across KC subtypes.  
        \textbf{G} Scatter plot of the fraction of the proximal state versus \textit{rut}'s mean transcriptional level for each E--P pair and across KC subtypes. Error bars in panels E-G represent 95\% confidence intervals from 1000 bootstrap resamples. Shaded areas in panels E-G represent confidence levels.
}
\label{fig:fig4}
\end{figure}

\noindent Our previous analyses comparing KCs and non-KCs showed that E--P proximal states are enriched in transcriptionally active cells. However, it remains unclear whether E--P proximity merely distinguishes transcriptionally active from inactive cells or whether it quantitatively predicts transcriptional output. If the latter is true, neuronal subtypes exhibiting higher \textit{rut} transcription should display a corresponding decrease in E--P distance. We tested this hypothesis by analyzing E--P proximity across the three Kenyon cell subtypes, $\alpha\beta$-KCs (49\%), $\alpha'\beta'$-KCs (18\%), and $\gamma$-KCs (33\%) \cite{Aso2009-zk}, which display distinct transcriptional and accessibility patterns. The three KC subtypes transcribe \textit{rut}, with $\alpha\beta$-KCs and $\alpha'\beta'$-KCs showing the highest and $\gamma$-KCs the lowest transcriptional levels (Figs.~\hyperref[fig:fig4]{4A}, \hyperref[fig:figS4]{S4A}). The accessibility of $P_{rut}$ is similar across KC subtypes and non-KC neurons (Fig.~\hyperref[fig:figS4]{S4B}). In contrast, $e_1$ and $e_2$ enhancer accessibility differs considerably among KC subtypes: high in $\alpha\beta$-KCs, intermediate in $\gamma$-KCs, and low in $\alpha'\beta'$-KCs (Fig. \hyperref[fig:fig4]{~4B}). These differences allowed us to test whether accessibility instructs E--P proximity and transcriptional output within KCs.\\

\noindent To investigate whether enhancer accessibility instructs the formation of E--P proximal states, we leveraged the tunable specificity of the UAS-GAL4 system to drive GFP expression in different KC subtypes using subtype-specific GAL4 drivers: c739$>$GAL4 for $\alpha\beta$-KCs, c305a$>$GAL4 for $\alpha'\beta'$-KCs, and H24$>$GAL4 for $\gamma$-KCs (Fig. \hyperref[fig:figS4]{~S4C}) (\hyperref[sec:methods]{Methods}). Each GAL4 stock was crossed with a UAS-nls-GFP line to drive nls-GFP expression in a single KC subtype at a time. Offspring from these crosses were used to perform Hi-M, enabling us to reconstruct 3D chromatin organization in each KC subtype (Fig. \hyperref[fig:fig4]{~4C}).\\

\noindent The fraction of proximal states involving the promoter and KC-specific enhancers is consistently higher in all KC subtypes than in non-KCs (Fig.~\hyperref[fig:fig4]{4D}). However, the high accessibility of $\alpha\beta$-KCs is not accompanied by increased E--P proximal states in this subtype, hinting at a decoupling between accessibility and spatial proximity. Instead, elevated fractions are observed in $\alpha'\beta'$- and $\gamma$-KCs, despite their lower enhancer accessibility compared to $\alpha\beta$-KCs. These observations are consistent with direct measurements of absolute E--P proximity frequencies (Fig.~\hyperref[fig:figS4]{S4D}).\\

\noindent To quantitatively assess whether accessibility correlates with E--P proximity, we integrated pseudo-bulk ATAC-seq signals for each enhancer barcode across KC subtypes and non-KCs and computed their median pairwise distance to the \textit{rut} promoter (Fig. \hyperref[fig:fig4]{4E}). This analysis reveals, in average, the higher the enhancer is accessible the closer it is to \textit{rut}'s promoter (Pearson coefficient $\approx-0.31$). We note, however, substantial deviations from this trend: many enhancers display high accessibility but moderate proximity to the promoter,  while other enhancers with low accessibility are unexpectedly close to the promoter. Importantly, most of this correlation is driven by differences between KC and non-KCs, and when this analysis is restricted to KC subtypes, higher enhancer accessibility is instead linked to higher median distances to the promoter (grey curve, Pearson coefficient $\approx0.54$, Fig.~\hyperref[fig:fig4]{4E}). For example, the \textit{rut} promoter interacts more frequently with the $E_1$ enhancer in $\alpha'\beta'$ KCs than in $\gamma$ or $\alpha\beta$ KCs, despite displaying higher chromatin accessibility. Similar trends were observed when the analysis was performed for all barcodes in the \textit{rut} locus (Fig.~\hyperref[fig:figS4]{S4E}). Surprisingly, these results indicate that enhancer accessibility is not a good predictor of E--P proximity changes between KC subtypes at the \textit{rut} locus.\\

\noindent We next asked whether E--P proximity itself correlates with transcriptional output. For this, we compared \textit{rut} expression levels derived from scRNA-seq data with E--P median pairwise distances. When comparing non-transcribing cells (non-KCs) with transcribing cells (KCs), we observe a strong correlation between \textit{rut} transcription and E--P median distance (Pearson coefficient $\approx0.87$, red dashed line, Figs.~\hyperref[fig:fig4]{4F}, ~\hyperref[fig:figS4]{S4D}). Unexpectedly, the picture is more complex when comparing different KC subtypes against each other, where the correlation between E--P distance and transcription is considerably weaker (Pearson coefficient $\approx 0.26$, black dashed line, Fig.~\hyperref[fig:fig4]{4F}). In addition, we note that while the transcriptional levels of $\alpha\beta$-KCs and $\alpha'\beta'$-KCs are similar, their E--P distances display large variations. Moreover, $\gamma$-KCs display the lowest transcriptional levels of the three subtypes, yet their E--P distances are similar to those of $\alpha'\beta'$-KCs, and within the spread of E--P distances of $\alpha\beta$-KCs. \\

\noindent To further consolidate these results, we plotted the normalized \textit{rut} expression as a function of the fraction of proximal state for all E--P pairs. While a positive correlation is observed when non-KCs and KC subtypes are both considered (Pearson coefficient $\approx 0.78$, red dashed line, Fig.~\hyperref[fig:fig4]{4G}), this correlation breaks down when one considers only KC subtypes. In this case, \textit{rut}'s transcriptional levels and fraction of alleles in the proximal state are in fact anti-correlated (Pearson coefficient  $\approx -0.83$, black dashed line, Fig.~\hyperref[fig:fig4]{4G}). This decoupling between transcriptional output and E--P proximity is unlikely to arise from a simple enhancer-switching mechanism, whereby distinct KC subtypes rely on different enhancers to drive \textit{rut} expression, since both candidate enhancers are accessible and active across all three KC subtypes. Thus, these results indicate that E--P distance or fraction of proximal state alone are poor predictors of transcriptional output among active cell types.\\

\noindent Taken together, our results show that E--P proximity distinguishes active from inactive neuronal populations, but this relationship breaks down within KCs, where variations in E--P proximity and enhancer accessibility are largely uncoupled from transcriptional levels. These findings support a model in which E--P proximity defines a permissive structural state, while transcriptional output is tuned by other mechanisms, including cell-type-specific transcription-factor/co-factors, promoter competence, and/or downstream transcriptional regulation.\\

\section*{Discussion}
In this study, we implemented Hi-M in the adult \textit{Drosophila} brain to visualize 3D chromatin organization in distinct neuronal cell-types at single-cell resolution. This approach enabled us to directly compare E--P proximity across neuronal populations displaying distinct transcriptional states inferred from cell-type-specific gene expression programs. Our results provide insight into several unresolved questions regarding the relationship between chromatin architecture, enhancer activity, and transcriptional regulation in the adult fly brain.\\

\noindent First, the single-cell nature of our measurements enabled us to resolve two distinct E--P conformational states that would be obscured in population-averaged analyses: a proximal state in which enhancers and promoters are in close spatial proximity, and a distal state in which they remain separated. While dynamic E--P topological states have been suggested in previous studies in engineered experimental systems \cite{Chen2018-rn}, our data provide direct structural evidence for the coexistence of proximal and distal configurations at an endogenous locus in an intact, fully differentiated tissue. While our measurements do not provide information on the kinetics or duration of these interactions, the relative abundance of proximal states provides an estimate of the probability with which enhancers occupy promoter-proximal conformations.\\

\noindent These probabilities are considerably higher than previously measured \cite{Espinola2021-mw, Mateo2019-wg, Benabdallah2019}, and show that \textit{rut}'s promoter is frequently ($\approx50\%$ of the time) found in close proximity to its enhancers specifically in transcriptionally-active cells. These observations provide strong evidence that 3D chromatin architecture contributes to enhancer function in differentiated neurons. While our data do not provide direct information on the lifetime of E--P interactions, the high frequency of promoter-proximal states strongly supports models in which recurrent E--P encounters play an active role in transcriptional activation. Nevertheless, our measurements cannot fully discriminate between stable-loop and hit-and-run mechanisms \cite{Karr2022-af,Yang2024-jx}, as both could in principle could generate frequent promoter-proximal states, depend on the kinetics of transcription. Resolving this distinction will require direct, highly accurate measurements of E--P interaction dynamics and transcription, which are currently challenging in living organisms \cite{Chen2023-aq}.\\

\noindent Second, current models propose that cell-type-specific chromatin accessibility changes at enhancer sequences are achieved and maintained through the recruitment of cell-type-specific TFs, which locally displace nucleosomes to enable the recruitment of co-activators and chromatin regulators \cite{Klemm2019-bb}. At the \textit{rut} locus, these accessibility changes are accompanied by increased frequencies of proximal E--P states in KC neurons relative to non-KC neurons. Thus, enhancer accessibility is associated with a redistribution of chromatin conformations toward spatially proximal E--P states. Interestingly, whereas activation of the \textit{rut} locus is associated with increased frequencies of proximal E--P states, other systems have reported transcription-associated increases in apparent E--P distances following enhancer activation and chromatin opening, suggesting that the structural consequences of enhancer activation may be locus- and context-dependent \cite{Gomez-Acuna2024-wd}. Together, these observations support a model in which chromatin opening at the \textit{rut} locus is associated with an increased probability of promoter-proximal configurations and transcriptional activation. In this model, E--P proximity appears to contribute to the establishment of a transcriptionally active state, although its relationship to quantitative transcriptional output may depend on additional regulatory mechanisms.\\

\noindent These observations contrast with previous studies in early \textit{Drosophila} embryogenesis showing that E--P proximity and transcriptional activation are largely decoupled \cite{Espinola2021-mw, Ing-Simmons2021}. Instead, our results are consistent with comparisons of E--P contact frequencies between neurons and muscle cells during late embryogenesis \cite{Pollex2024}. Importantly, our results reveal that distinct E--P architectures are observed between different post-mitotic neuronal cell-types within the adult brain, extending previous work and suggesting that enhancer–promoter proximity remains tightly associated with transitions between inactive and active transcriptional states at late stages of neuronal differentiation.\\

\noindent Third, comparisons between KC subtypes revealed that quantitative differences in chromatin accessibility are not associated with corresponding changes in either E--P proximity or \textit{rut} transcription. These observations suggest that, within active neuronal populations, chromatin accessibility is better interpreted as a marker of regulatory competence than as a direct quantitative predictor of enhancer activity, promoter engagement, or transcriptional output. This interpretation is consistent with a growing body of work showing that accessible chromatin is widely used to identify candidate enhancers, yet accessibility alone is often insufficient to explain functional enhancer activity in specific cellular contexts \cite{Catarino2018-ux}. Instead, the action of KC enhancers in different KC subtypes may depend on additional regulatory layers including TF occupancy, cofactor recruitment, and promoter responsiveness, which may in turn regulate promoter activation dynamics and/or transcriptional bursting behavior \cite{Spitz2012-gq, Shlyueva2014-qy,Fukaya2016-ls}. Such mechanisms may reflect the partially shared yet subtype-specific transcription factor networks that define the regulatory identities and neuronal functions of KC subtypes \cite{Janssens2022}, which could modulate \textit{rut} expression within a shared permissive chromatin landscape without requiring systematic changes in enhancer–promoter encounter frequencies. Together, our observations suggest that chromatin accessibility and E--P proximity at the \textit{rut} locus contribute to establishing a transcriptionally permissive state, but are not quantitatively instructive for transcriptional output across KC subtypes. Comparisons among KC populations revealed that quantitative differences in transcription among active neuronal populations are likely specified by additional regulatory mechanisms operating downstream or independently of chromatin proximity.\\

\noindent Several limitations should be considered when interpreting our results. First, our study focuses on the \textit{rut} locus, which represents a biologically relevant model for neuronal enhancer regulation, but does not necessarily establish universal principles for enhancer–promoter organization across all neuronal genes. Second, cell-type assignment relies on GAL4 drivers combined with immuno-Hi-M segmentation, and therefore depends on the specificity and completeness of available genetic labels. Third, because Hi-M measurements are performed in fixed cells, our analysis resolves the frequencies of chromatin conformational states across populations of cells, but cannot directly measure the kinetics of transitions between proximal and distal states or their temporal relationship to transcriptional bursting. Finally, transcriptional states were inferred from previously published single-cell transcriptomic datasets rather than measured simultaneously at individual cells together with chromatin architecture, which limits our ability to directly relate enhancer–promoter proximity to instantaneous transcriptional output.\\

\noindent Last but not least, this work establishes single-cell 3D chromatin organization imaging in defined adult fly neuronal populations while preserving anatomical context. More broadly, extending spatial genome imaging approaches to other intact adult tissues will enable future studies addressing how chromatin architecture contributes to cell identity, physiological adaptation, aging, regeneration, and disease across complex organs and developmental contexts.

\section*{Methods and Materials}
\label{sec:methods}

\subsection*{\textit{Drosophila} strains and oligopaint libraries}
\noindent Fly stocks were maintained on standard cornmeal–yeast medium at 25°C. The following lines were used: UAS-GFP.nls (Bloomington \#4775), OK107$>$GAL4 (Bloomington \#854) for all Kenyon cells (KCs), c739$>$GAL4 (Bloomington \#64305) for $\alpha\beta$ KCs, c305a$>$GAL4 (Bloomington \#30829) for $\alpha'\beta'$ KCs, and H24$>$GAL4 (gift from Jean-Maurice Dura) for $\gamma$ KCs. Homozygous GAL4 males were crossed with homozygous UAS-GFP.nls virgin females, and F1 progeny were used for all experiments.\\

\noindent The oligopaint library used in this study was previously described and validated in \cite{Bardou2026-iw}. Briefly, the library was designed using the Oligopaints database (\url{http://genetics.med.harvard.edu/oligopaints}) using 35/45mer sequences homologous to the \textit{Drosophila melanogaster} genome (dm6). Hi-M probe sets were designed against the \textit{rut} locus (chrX:14785864-14876400, dm6) according to the method previously described \cite{Cardozo_Gizzi2020-yx}. The library contained 25 barcodes, each approximately 3 kb in length. The full list of barcode and probe sequences is available in Ref. \cite{Bardou2026-iw}.\\

\subsection*{Sample preparation and Hi-M acquisitions}
\noindent Sample preparation, DNA-FISH labeling, and Hi-M acquisitions were as previously described \cite{Bardou2026-iw}. Briefly, adult flies (2–5 days old) were fixed in 4\% paraformaldehyde (PFA), brains were dissected, cryoprotected in 30\% sucrose, embedded in OCT, and cryosectioned into 10$\mu$m-thick sections. DNA-FISH was performed using the \textit{rut} oligopaint library described above.\\

\noindent DNA-FISH was performed as previously described \cite{Messina2024.09.18.613689}. Briefly, slides were treated with RNase A (200 \textmu g/ml in PBS) for 1 hour at room temperature (RT), followed by incubation in sodium citrate buffer (10 mM citrate, 0.05\% Tween, pH 6.0) for 5 minutes at RT, then 25 minutes at 80\textdegree C. After washing with 2xSSC, slides were incubated in 50\% formamide wash buffer for 2 hours at RT. They were then hybridized with 2 \textmu L of 5 to 10 \textmu g/\textmu L library and 1 \textmu L of 100 \textmu M of the fiducial library in hybridization buffer (50\% formamide, 10\% dextran sulfate, 2xSSC, 0.5 mg/mL Salmon Sperm DNA) for 3 hours at 45\textdegree C, followed by heat-shock at 85\textdegree C for 5 minutes. The slides were incubated overnight at 37\textdegree C in a humidity-controlled environment. The next day, slides were washed sequentially in formamide buffers (50\%, 40\%, 30\%, 20\%, 10\%) diluted in 2xSSC under agitation. Post-fixation was performed with 4\% PFA for 10 minutes at RT, and slides were stored in 2xSSC at 4\textdegree C until imaging.\\

\noindent As DNA-FISH labeling causes a loss of the GFP signal, immunolabeling was required to identify KCs and their subtypes. After DNA-FISH labeling, GFP was detected using a chicken polyclonal anti-GFP antibody (Invitrogen, catalog no. A10262; 1:200) and a goat anti-chicken secondary antibody conjugated to Alexa Fluor 647 (Invitrogen, catalog no. A21449; 1:500), as previously described \cite{Bardou2026-iw}.\\

\noindent Hi-M acquisitions were performed by sequential hybridization and imaging of the 25 readout probes of the \textit{rut} oligopaint library, together with a fiducial probe used for image registration. For each experiment, image stacks were acquired from 10--20 regions of interest (ROIs) of $200$\textmu m$\times200$\textmu m, with a z-step size of $250nm$ over a total range of $17.5$\textmu m. DAPI, the fiducial signal, and GFP immunolabeling were acquired before sequential imaging of the 25 Hi-M barcodes. Complete details of the sample preparation, sequential hybridization, imaging, and image acquisition procedures are provided elsewhere \cite{Bardou2026-iw}.\\

\noindent The analysis of Hi-M datasets started by the deconvolution of raw TIFF images using Huygens Professional 21.04 (Scientific Volume Imaging, \url{https://svi.nl}). The following analyses steps were performed using pyHiM (\url{https://pyhim.readthedocs.io/en/latest/}) release 0.7 \cite{Devos2024-uw} and were described in more details elsewhere \cite{Bardou2026-iw}. Briefly, these steps involved: (1) registration of sequential cycles using \textit{register\_global} and \textit{register\_local}; (2) segmentation of oligopaint masks using \textit{mask\_3d}; (3) segmentation and localization of DNA-FISH spots using \textit{localize\_3d}; (4) tracing using \textit{build\_trace}; (5) KC subtypes for individual crosses were assigned based on GFP immunofluorescence. Segmentation and classification were performed using Ilastik (random forest classifier), followed by the \textit{trace\_assign.py} script from \href{https://traceratops.readthedocs.io/}{traceratops} to assign chromatin traces to specific neuronal populations.\\

\subsection*{Post-processing of raw traces}
\noindent Raw traces were post-processed using scripts from \href{https://traceratops.readthedocs.io/}{traceratops}, as follows: (1) raw trace tables from different replicates \cite{Bardou2026-iw} were merged using \textit{trace\_merge.py} see Table \ref{merged-tracetables-filenames}; (2)  \textit{trace\_filter.py} was used to remove traces with less than three barcodes, and barcodes that were present more than once in a trace; (3) traces were split using \textit{trace\_split.py}. The standard deviation threshold was set to $1.0$ and the number of clusters to $2.0$; (4) missing barcodes were imputed using the following procedure: if a barcode was absent but its two adjacent barcodes were detected, its spatial position was estimated as the midpoint between the neighboring loci, as previously described \cite{Messina2024.09.18.613689}; (5) To further account for localization uncertainty inherent to multi-locus imaging, we applied a post-processing correction strategy termed LEAR (Loci Enabled Advanced Resolution) \cite{Bohrer2025-nz}. Briefly, LEAR takes advantage of the fact that the imaged loci have known identities and relative positions along the traced region. Given an estimate of the localization error in each imaging dimension, the method estimates how much of the observed pairwise displacement variance is expected to come from imaging error, and then adjusts the observed barcode positions so that the traces better match this corrected variance. Here, we used a conservative estimate of $30nm$ for the x/y localization error and accounted for the larger uncertainty along z, which was estimated empirically to be approximately $100nm$; (6) Finally, construction of pairwise-distance and proximity matrices was performed using \textit{trace\_to\_matrix.py}. For this, we used a distance threshold of $150nm$, a value that optimizes the correlation between Hi-M and Hi-C datasets (Figs.~\hyperref[fig:fig1]{1G} and ~\hyperref[fig:figS2]{S2B}). Post-processed trace tables used in this manuscript are listed in Table \ref{table-tracetables-filenames}. These tables can be downloaded from our \href{https://osf.io/vbpuy/?view_only=9968c35f4d9242b4ad4b329f64ad48cc}{Open Science Foundation (OSF) Repository}. \\

\begin{table}[htbp]
\centering
\begin{tabular}{|c|c|c|c|c|}

\hline
\textbf{\makecell{Raw Trace file}} &
\textbf{Label} &
\textbf{Post-processed table}\\
\hline
{\fontsize{8pt}{9.5pt}\selectfont \makecell{
traces\_KCs\_rep1.ecsv, traces\_KCs\_rep2.ecsv \\ traces\_KC\_AB\_rep1.ecsv, traces\_KC\_AB\_rep2.ecsv\\ traces\_KC\_ABp\_rep1.ecsv, traces\_KC\_ABp\_rep2.ecsv, traces\_KC\_ABp\_rep3.ecsv\\ traces\_KC\_G\_rep1.ecsv, traces\_KC\_G\_rep2.ecsv, traces\_KC\_G\_rep3.ecsv\\ traces\_nonKC\_rep1.ecsv, traces\_nonKC\_rep2.ecsv\\ traces\_nonKCs\_KCs\_merged.ecsv}} & \makecell{non-KCs\\ KCs} & traces\_flybrain.ecsv\\ 
\hline
{\fontsize{8pt}{9.5pt}\selectfont \makecell{traces\_nonKC\_rep1.ecsv, traces\_nonKC\_rep2.ecsv}} & \makecell{non-KCs} & traces\_non\_KCs.ecsv\\ 
\hline
{\fontsize{8pt}{9.5pt}\selectfont \makecell{traces\_KCs\_rep1.ecsv, traces\_KCs\_rep2.ecsv \\ traces\_KC\_AB\_rep1.ecsv, traces\_KC\_AB\_rep2.ecsv\\ traces\_KC\_ABp\_rep1.ecsv, traces\_KC\_ABp\_rep2.ecsv, traces\_KC\_ABp\_rep3.ecsv\\ traces\_KC\_G\_rep1.ecsv, traces\_KC\_G\_rep2.ecsv, traces\_KC\_G\_rep3.ecsv\\}} & \makecell{KCs} & traces\_KCs.ecsv\\ 
\hline
{\fontsize{8pt}{9.5pt}\selectfont \makecell{traces\_KC\_AB\_rep1.ecsv, traces\_KC\_AB\_rep2.ecsv}} & \makecell{KC-$\alpha\beta$} & traces\_KC\_AB.ecsv\\ 
\hline
{\fontsize{8pt}{9.5pt}\selectfont \makecell{traces\_KC\_ABp\_rep1.ecsv, traces\_KC\_ABp\_rep2.ecsv, traces\_KC\_ABp\_rep3.ecsv}} & \makecell{KC-$\alpha'\beta'$} & traces\_KC\_ABp.ecsv\\ 
\hline
{\fontsize{8pt}{9.5pt}\selectfont \makecell{traces\_KC\_G\_rep1.ecsv, traces\_KC\_G\_rep2.ecsv, traces\_KC\_G\_rep3.ecsv\\}} & \makecell{KC-$\gamma$} & traces\_KC\_G.ecsv\\ 
\hline
\end{tabular}
\caption{\label{tab3} Generation of merged trace files from the original raw trace files. Each post-processed trace file was generated by merging the corresponding
raw trace files from the indicated replicates.}
\label{merged-tracetables-filenames}
\end{table}
\FloatBarrier

\begin{table}[htbp]
\centering
\begin{tabular}{|c|c|c|c|c|}
\hline
\textbf{\makecell{Trace table}} &
\textbf{Cell-type} &
\textbf{N unique traces} &
\textbf{ROIs} &
\textbf{Figures used} \\
\hline
traces\_flybrain.ecsv & KC and non-KC & 332,537 & 211 & 1, S1, S2 \\
\hline
traces\_KCs.ecsv & KC & 28,770  & 56 & 1, 2, 3, S1, S2, S3\\
\hline
traces\_non\_KC.ecsv & non-KC & 31,505  & 11 & 2, 3, 4, S3, S4\\
\hline
traces\_KC\_AB.ecsv & KC-$\alpha\beta$ & 6,632  & 26 & 1, 4, S1, S2, S4\\
\hline
traces\_KC\_ABp.ecsv & KC-$\alpha'\beta'$ & 5,522  & 40 & 1, 4, S1, S2, S4 \\
\hline
traces\_KC\_G.ecsv & KC-$\gamma$ & 6,982  & 33 & 1, 4, S1, S2, S4\\
\hline
\end{tabular}
\caption{\label{tab4} Post-processed trace tables. Unique traces: number of unique traces with more than two barcodes. ROI: region of interest}
\label{table-tracetables-filenames}
\end{table}
\FloatBarrier

\subsection*{Insulation score analysis from Hi-M data}
\noindent Insulation scores were computed from Hi-M pairwise distance matrices following previously described approaches \cite{Messina2023-tw}. Briefly, a square window of size $n \times n$ was slid along the diagonal of the median pairwise distance matrix, and the values within each window were aggregated to quantify local insulation strength. To probe insulation across multiple genomic scales, a domainogram was generated by repeating this calculation over a range of window sizes, ranging from 3 kb to 18 kb (1–6 bins). The resulting matrix was subsequently smoothed using spatial interpolation to enhance the visualization of domain boundaries.\\

\subsection*{Normalization of Hi-M maps}
\noindent To account for the dependence of physical distance on genomic separation, we first computed the mean physical distance ($\mu$m) as a function of genomic distance (kb) using Hi-M data (Fig.~\hyperref[fig:figS2]{S2A}). This relationship was used to estimate the expected physical distance for each pair of loci given their genomic separation. Pairwise distances were then normalized by dividing the observed (O) physical distances by the expected (E) values derived from this mean trend. A log-transformed observed-over-expected metric was subsequently computed as $\log_2(\mathrm{O}/\mathrm{E})$. The resulting distance normalized matrices were used to highlight deviations from average distance.\\

\subsection*{Volcano plot based on HiM}
\noindent Volcano plots from Hi-M matrices were computed for two conditions. First, the differential median PWD matrix was calculated by determining the ratio of each pair of bins between the two conditions. Then, p-values were derived using the Wilcoxon rank-sum test, comparing the single-cell distance distributions for each pair of bins. Finally, the log2 values of the differential matrix were plotted against the -log10(p-values) from the Wilcoxon test.\\

\subsection*{PWD distribution analysis}
\label{sec:pwd_analysis}

\noindent To quantitatively analyze the spatial proximity distributions between the \textit{rut} promoter and other genomic loci within the \textit{rut} locus, we employed a two-state model previously described in Refs. \cite{Remini2024-aa,Remini2025-jcp,Polovnikov2023-prx}. \noindent The distance probability distribution \(P_{ij}(r)\) between genomic loci \(i\) and \(j\) in an ideal polymer follows a Gaussian form that depends on the mean-squared radius \(R^2 = \langle \vec{r}^2 \rangle\) as its sole parameter. For a random walk, the characteristic radius scales as \(R \sim |i-j|^{1/2}\). However, this model provides only a reasonable first approximation to the complex phases of polymeric matter. A well-known result of polymer physics is that a single test chain immersed in a melt of other chains behaves as an ideal polymer due to the screening of self-avoidance interactions by surrounding polymers \cite{Degennes1979}. It has been recently shown that the introduction of suitable monomer-monomer harmonic pairwise interactions in a bead-and-spring polymer model leads to distance distributions described by fractional Brownian motion (fBm) \cite{Polovnikov2018-sm}, with characteristic radius scaling \(R \sim |i-j|^\nu\), where the exponent can be tuned in the interval \(1/3 \leq \nu \leq 1/2\). This range encompasses the crumpled globule phase exponent (\(\nu=1/3\)) computed for chromatin \textit{via} Hi-C experiments \cite{Lieberman2009-sci}.\\

\noindent Despite the wide-ranging applicability of single Gaussian models in polymer physics, recent analyses of experimental sequential DNA-FISH data \cite{Remini2024-aa,Remini2025-jcp} and Hi-C measurements \cite{Polovnikov2023-prx} of chromatin architecture reveal significant deviations from a single phase behavior. These deviations are particularly pronounced at short genomic distances ($|i-j| \leq 500$ kb), where the pairwise interaction we investigate in this work occurs. These findings suggest the coexistence of distinct chromatin conformational states, each with unique mechanical properties and spatial organization. Consequently, the observed distribution of pairwise distances (PWDs) between genomic segments \(i\) and \(j\), denoted by \(P_{ij}(r)\), was fitted by a linear combination of two distributions representing Proximal (P) and Non Proximal (NP) enhancer-promoter states.
\begin{equation}
P_{ij}(r) = f_{P}\, g(r; R_{P}) + (1 - f_P)\, g(r; R_{NP}),
\label{eq:two_state_pwd}
\end{equation}

\noindent where \( f_P \) and \( f_{NP} = 1 - f_P \) represent the fractions of enhancer-promoter pairs that exist in the proximal and non-proximal states, respectively. The distribution \( g(r; R) \) for distance \( r \) and characteristic radius \( R \) is defined as:

\begin{equation}
g(r; R) = 4 \pi r^2 \left(\frac{3}{2\pi R^2}\right)^{3/2} \exp\left(-\frac{3r^2}{2R^2}\right),
\label{eq:gaussian_pwd}
\end{equation}

\noindent where the term \(4\pi r^2\) arises from the Jacobian transformation to spherical coordinates. This functional form is normalized and encompasses a broad range of polymer models, from ideal chains to crumpled globules.\\

\noindent We assume ergodicity within each chromatin state, allowing us to equate time averages along single-locus trajectories with ensemble averages over polymer configurations \cite{Khanna2019-tm}. Consequently, each chromatin state corresponds to a distinct ensemble of configurations. The proximal state (P) comprises configurations where the enhancer and promoter are brought into close spatial proximity through protein-mediated bridging interactions. In contrast, the non-proximal state (NP) consists of more extended configurations where enhancer–promoter contacts are disrupted. The two-phase model serves as a direct proxy for the dynamic equilibrium between these functionally distinct ensembles, enabling quantitative inference of changes in enhancer–promoter interaction prevalence across cell-types by fitting the parameter \(f_P\).\\

\noindent To ensure the biological relevance and stability of the fits, we imposed specific constraints on the characteristic radii of each state. The proximal state (\(R_{P}\)) was constrained to be less than 200 nm, while the non-proximal state (\(R_{NP}\)) was constrained to be greater than 200 nm. These thresholds were chosen based on previous experimental considerations regarding the spatial scales of enhancer-promoter interactions, as validated in Fig.~\hyperref[fig:figS2]{S2B}.\\

\noindent The parameters of the model, including the fraction of the proximal state (\(f_{P}\)) and the characteristic radii of the proximal and non-proximal states (\(R_{P}, R_{NP}\)), were determined using non-linear least squares optimization on the empirical probability density functions. The uncertainties on each parameter were obtained from the covariance matrix of the fits. The quality of the fits was assessed using the coefficient of determination (\(\mathcal{R}^2\)).\\
\begin{equation}
\mathcal{R}^{2} = 1 - \frac{\sum_{i=1}^{N_{\mathrm{bins}}} \left( p_i^{\mathrm{exp}} - p_i^{\mathrm{model}} \right)^2}{\sum_{i=1}^{N_{\mathrm{bins}}} \left( p_i^{\mathrm{exp}} - \bar{p}^{\mathrm{exp}} \right)^2},
\end{equation}
where \(p_i^{\mathrm{exp}}\) and \(p_i^{\mathrm{model}}\) denote the experimental and model-predicted probability densities in bin \(i\), respectively, and \(\bar{p}^{\mathrm{exp}}\) is the mean of the experimental probability densities. The resulting \(\mathcal{R}^2\) values (where \(\mathcal{R}^2 = 1\) indicates a perfect fit) were compared to those obtained from a single-Gaussian fit in order to quantify the improvement provided by the two-state description.\\

\noindent The fraction of the proximal state (\(f_{P}\)) for each pair of genomic regions was compared between Kenyon cells (KCs) and non-Kenyon cells (non-KCs) by calculating their respective uncertainties from the fitting procedure. The error bars presented in the manuscript represent the standard error derived from these uncertainties, facilitating a direct comparison of \(f_P\) values between cell-types. This comparison enables a statistical assessment of the differences in enhancer-promoter engagement dynamics.

\subsection*{Enhancer prediction}
\noindent Predicted enhancers along the \textit{rut} locus were identified using resources provided by the Janelia FlyLight Project Team (\url{https://flweb.janelia.org/cgi-bin/flew.cgi}).\\

\subsection*{Hi-C, ChIP-seq, ATAC-seq, RNA-seq and snRNA-seq data processing}
\noindent Hi-C data acquired in adult fly brains were downloaded in .mcool format from accession number GSE228095 \cite{Mohana2023}. CTCF and CP190 ChIP-seq acquired in third-instar larval CNSs were downloaded in bigwig format from accession number GSE146749 \cite{Kaushal2021}. Pseudo-bulk ATAC-seq acquired in different adult cell-types were downloaded from \url{https://flybrain.aertslab.org/downloads/bigwig_adult.html)} in bigwig format. When necessary, the dataset was lifted over from dm3 to dm6 coordinates using the LiftOver tool (\url{http://genome.ucsc.edu/cgi-bin/hgLiftOver}). Profiles of the genomic data were plotted along the \textit{rut} locus (chrX:14785864-14876400, dm6) using CoolBox (\url{https://gangcaolab.github.io/CoolBox/quick_start_API.html}). RNA-seq acquired in adult fly brains were downloaded from GSE37027 in bigwig format \cite{Henry2012}. Annotated single-cell RNA sequencing (snRNA-seq) data were downloaded from the SCope \cite{Janssens2022} and re-plotted using a custom Python script.\\

\section*{Data availability}
Data were deposited and can be downloaded from our publicly available \href{https://osf.io/vbpuy/?view_only=9968c35f4d9242b4ad4b329f64ad48cc}{Open Science Foundation (OSF) \textbf{Flybrain\_2025a} Repository}, under a CC-By Attribution 4.0 International license, and were assigned a permanent DOI: 10.12688/openreseurope.24301.1. Post-processed chromatin trace tables are compressed in: $traces\_processed.tar.gz$  within the same repository. The list of other published datasets used in this study is provided in Supplementary Table S1. The full list of barcode and probe sequences is available in Supplementary Data 1 and 2.\\

\section*{Code availability}
The code used to analyze chromatin tracing data is publicly available at: \href{https://github.com/pyHi-M/traceratops}{traceratops package}. The specific code to produce the figures is publicly available at: \href{https://github.com/NollmannLab-publications/messina_2026}{Messina\_2026 code}. The code for LEAR was made available previously \cite{Bohrer2025-nz}, and can be downloaded from: \href{https://github.com/CHB-Bohrer/LEAR}{LEAR code}.\\

\section*{Acknowledgments}
We thank Alistair Boettiger, Aleena Patel and Antoine Coulon for critical reading and comments on the manuscript. We thank Marion Bardou for her intellectual input, and Cedric Maurange, Giacomo Cavalli, and Enrico Carlon for their input and advice during the project. We thank Stein Aerts and Ibrahim Taskiran for their help in selecting the \textit{rut} locus. This project was funded by the European Union’s Horizon 2020 Research and Innovation Program (Grant ID 724429) (M.N.) and the French National Research Agency (ANR-23-CE12-0023-01) (M.N.). We acknowledge the Bettencourt-Schueller Foundation for their prize ‘Coup d'élan pour la recherche Française’. The CBS is a member of France-BioImaging, a national infrastructure supported by the French National Research Agency (ANR-10-INBS-04-01). O.M was supported by an FRM and Ligue Contre la Cancer PhD fellowships. This project was provided with computing (HPC) and storage resources by GENCI at CINES thanks to the grant allocation A0190316693 on the supercomputer Adastra GENOA/ MI250x partitions. L.R was supported by PhD fellowships from LabMUSE EpiGenMed within the I-Site MUSE (ANR-16-IDEX-0006) and from the ANR (TRANSLAxon project ANR-20-CE16-0025).\\

\section*{Author Contributions}
O.M., L.R., C.H.B. analyzed the data. O.M., L.R., X.D., M.N. wrote analysis software. J-B.F. built the microscope. O.M., L.R., C.H.B., A.P., M.N. interpreted the data. O.M., L.R. and M.N. wrote the manuscript. O.M., L.R., C.H.B., J-C.W., A.P. M.N. revised the manuscript. M.N. supervised the study and acquired funds.\\

\printbibliography

\section*{Figures}

\vspace{-20pt}

 \vspace{-300pt}

\section*{Supplementary Figures}

\setcounter{figure}{0}
\renewcommand{\thefigure}{S\arabic{figure}}
\vspace{-20pt}

\begin{figure}[H]
  \centering
  \includegraphics[height=0.8\textheight]{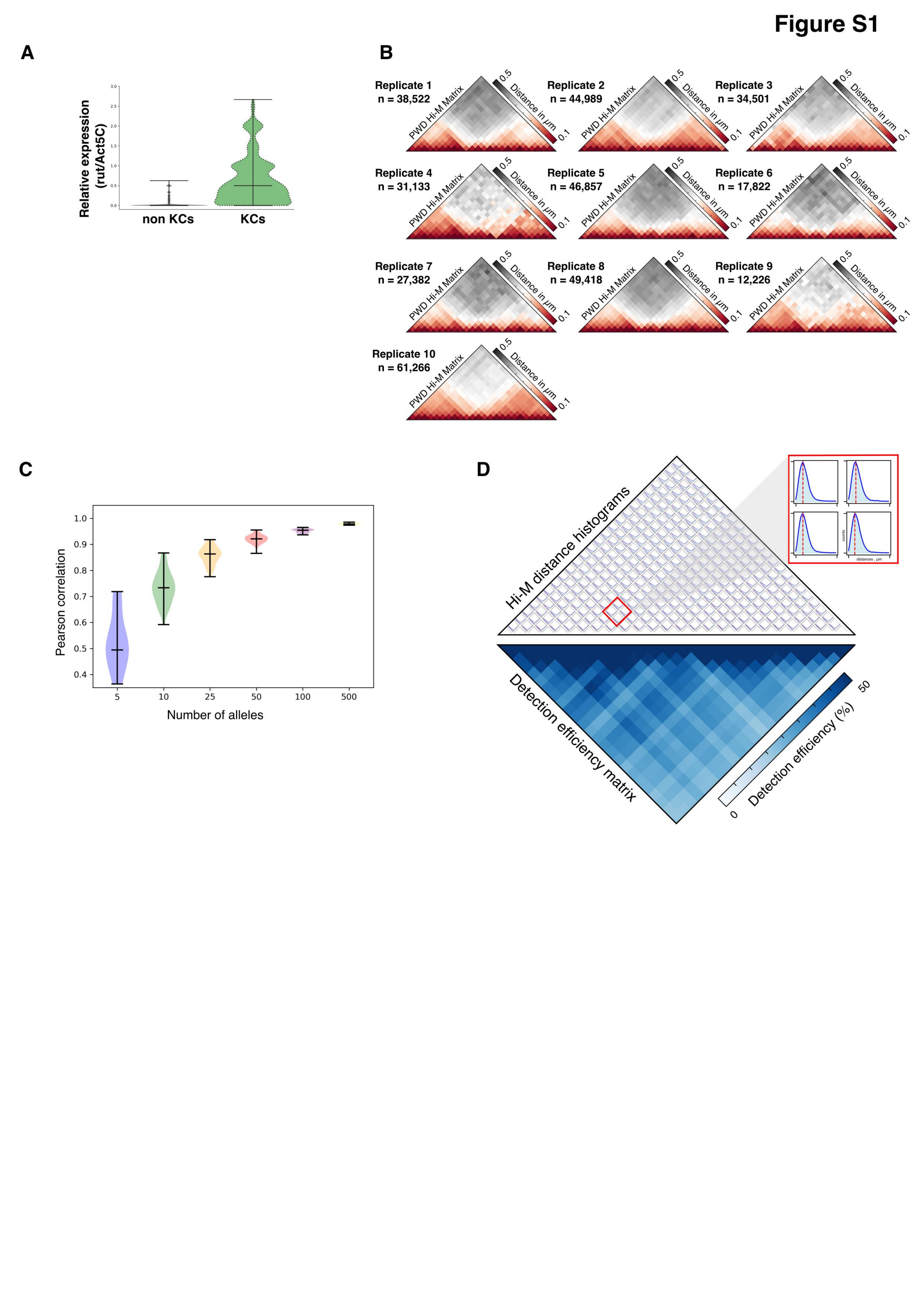}
  \captionsetup{
    width=\textwidth,
    justification=justified,
    singlelinecheck=false
  }
  \vspace{-200pt}
  \caption{%
    \textbf{A} Relative \textit{rut} expression levels derived from snRNA-seq experiments for non-Kenyon cells and Kenyon cells.
    \textbf{B} Median PWD matrices reconstructed for the \textit{rut} locus for each individual Hi-M experiment in the \textit{Drosophila} brain. n represents the number of traces used to construct each map. Hi-M maps were computed from raw traces. 
     \textbf{C} Violin plots showing the Pearson correlation between the ensemble PWD matrix derived from the full dataset and those obtained from subsets with varying numbers of single-cell data. Distributions were generated using bootstrapping with 50 cycles per condition.   
    \textbf{D} Histogram matrix of PWD for all barcode pairs obtained from the \textit{Drosophila} brain.  
  }
  \label{fig:figS1}
\end{figure}

\vspace{-60pt}
\begin{figure}[H]
  \centering
  \includegraphics[height=0.85\textheight]{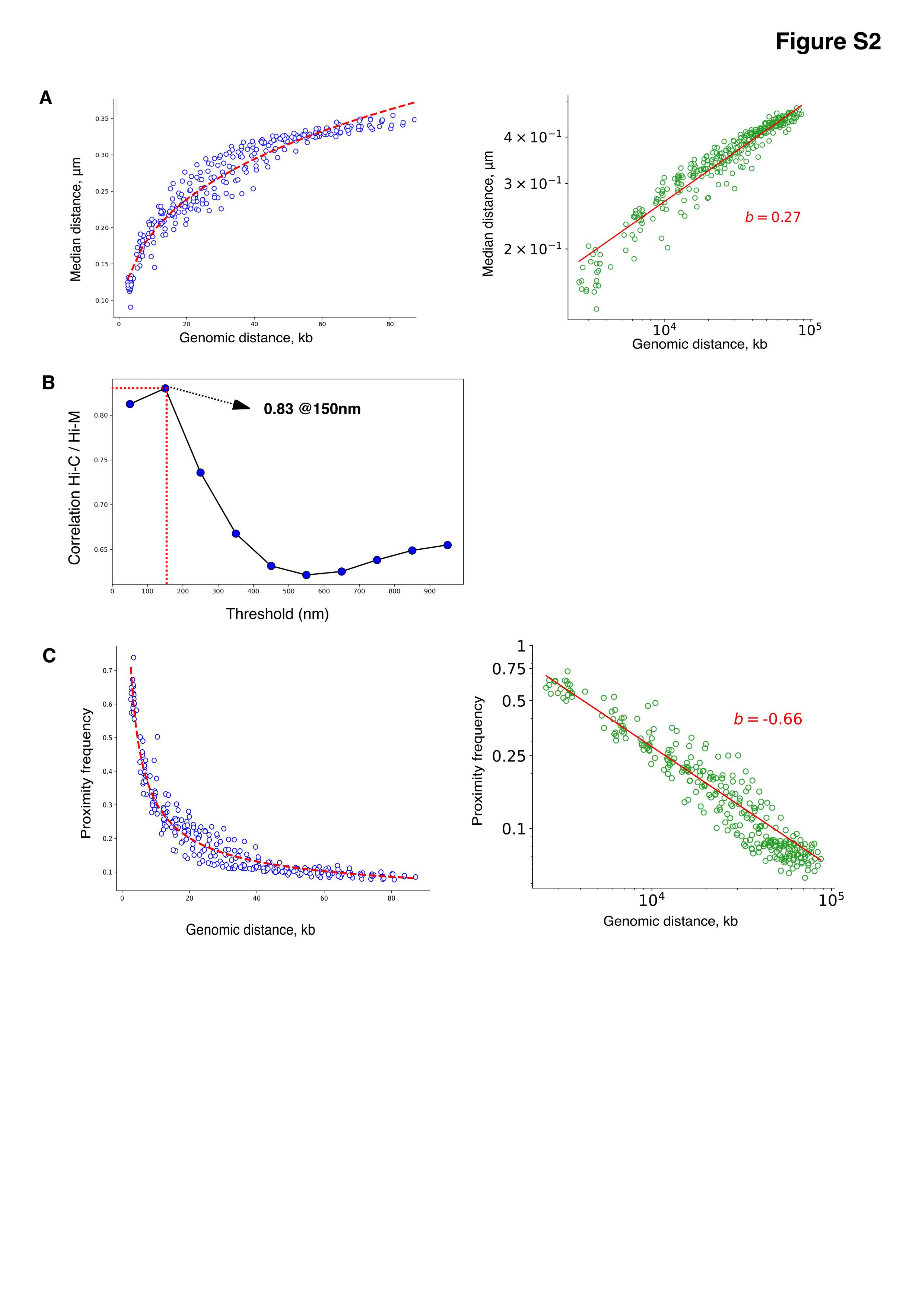}
  \captionsetup{
    width=\textwidth,
    justification=justified,
    singlelinecheck=false
  }
  \vspace{-150pt}
  \caption{\textbf{A} Left: Median PWD ($\mu$m) versus genomic distance (kb) from Hi-M data, shown on a linear scale. Red lines represent power-law fits to the experimental data. Right: The same data and fits displayed on a logarithmic scale. \textbf{B} Pearson correlation coefficient calculated between the interpolated Micro-C contact map and the Hi-M proximity frequency matrix across varying cut-off distances. The maximum correlation is observed at a threshold of $150$~nm, as indicated. \textbf{C} Left: Absolute proximity frequency versus genomic distance (kb) shown on a linear scale. Red lines represent power-law fits to the experimental data. Right: The same proximity frequency data displayed on a logarithmic scale.
  }
  \label{fig:figS2}
\end{figure}
\clearpage

\clearpage
\begin{figure}[H]
  \centering
  \includegraphics[height=0.85\textheight]{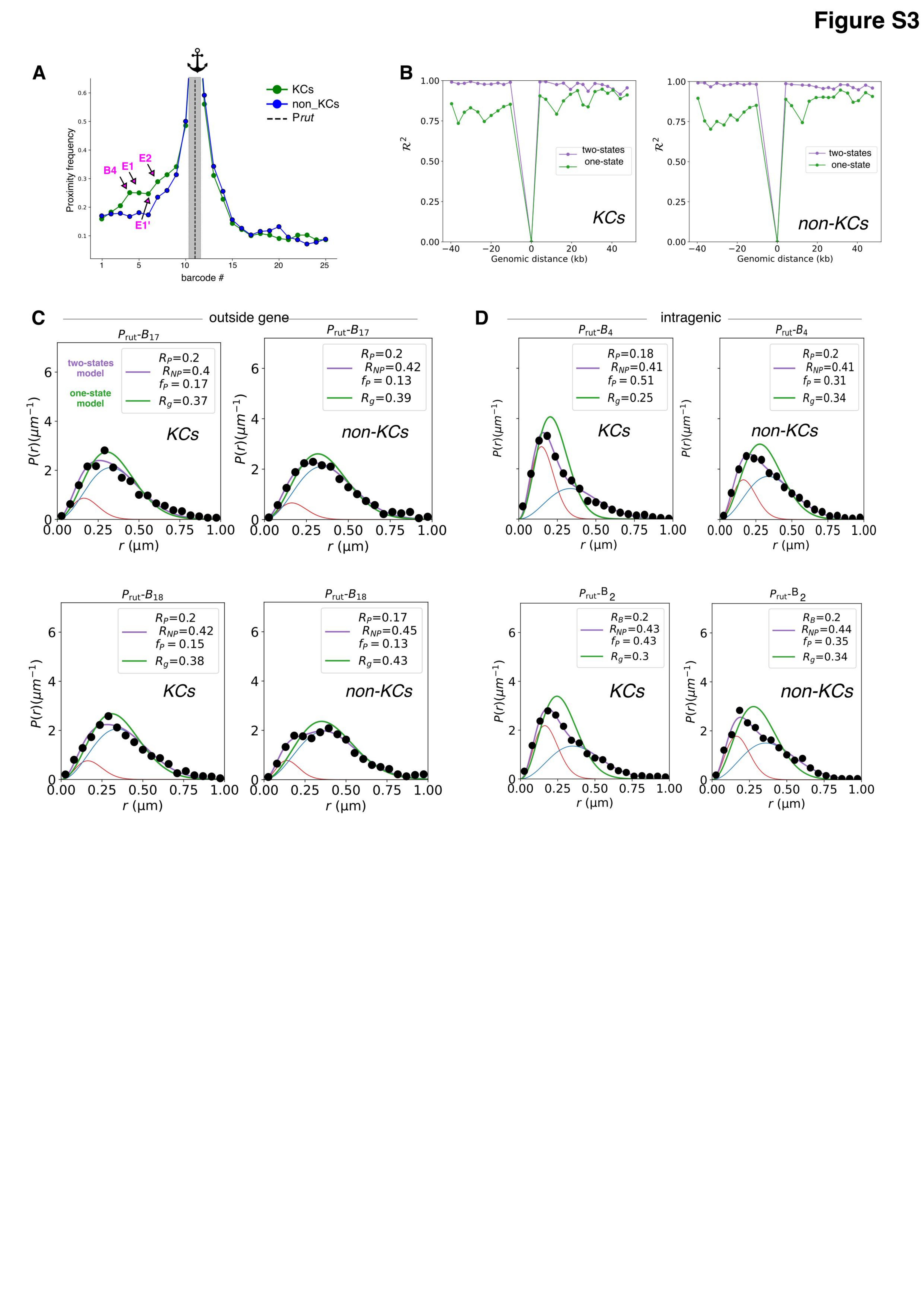}
  \captionsetup{
    width=\textwidth,
    justification=justified,
    singlelinecheck=false
  }
   \vspace{-200pt}
  \caption{%
    \textbf{A} Virtual 4M profiles anchored at the \textit{rut} promoter, for KCs (green lines) and non-KCs cells (blue line). \textbf{B} {Comparison of the Coefficient of Determination ($R^2$) for a two-state model (purple) versus a single-state Gaussian model (green) as a function of genomic distance in KCs and non-KCs.} \textbf{C} {Probability distributions of spatial distances $P(r)$ between $P_{rut}$ and outside-gene elements ($B_{17}$, top; $B_{18}$, bottom). Black dots represent experimental data, overlaid with fits for a one-state model ($R_g$) and a two-states model ($R_P$, $R_{NP}$).} \textbf{D} {Distance distributions $P(r)$ between $P_{rut}$ and intragenic elements ($B_4$, top; $B_2$, bottom). The two-states fit (purple) is further decomposed into its underlying bound ($R_B$, red) and unbound ($R_U$, blue) subpopulations.} 
  }
  \label{fig:figS3}
\end{figure}
\clearpage

\clearpage
\begin{figure}[H]
  \centering
  \includegraphics[height=0.85\textheight]{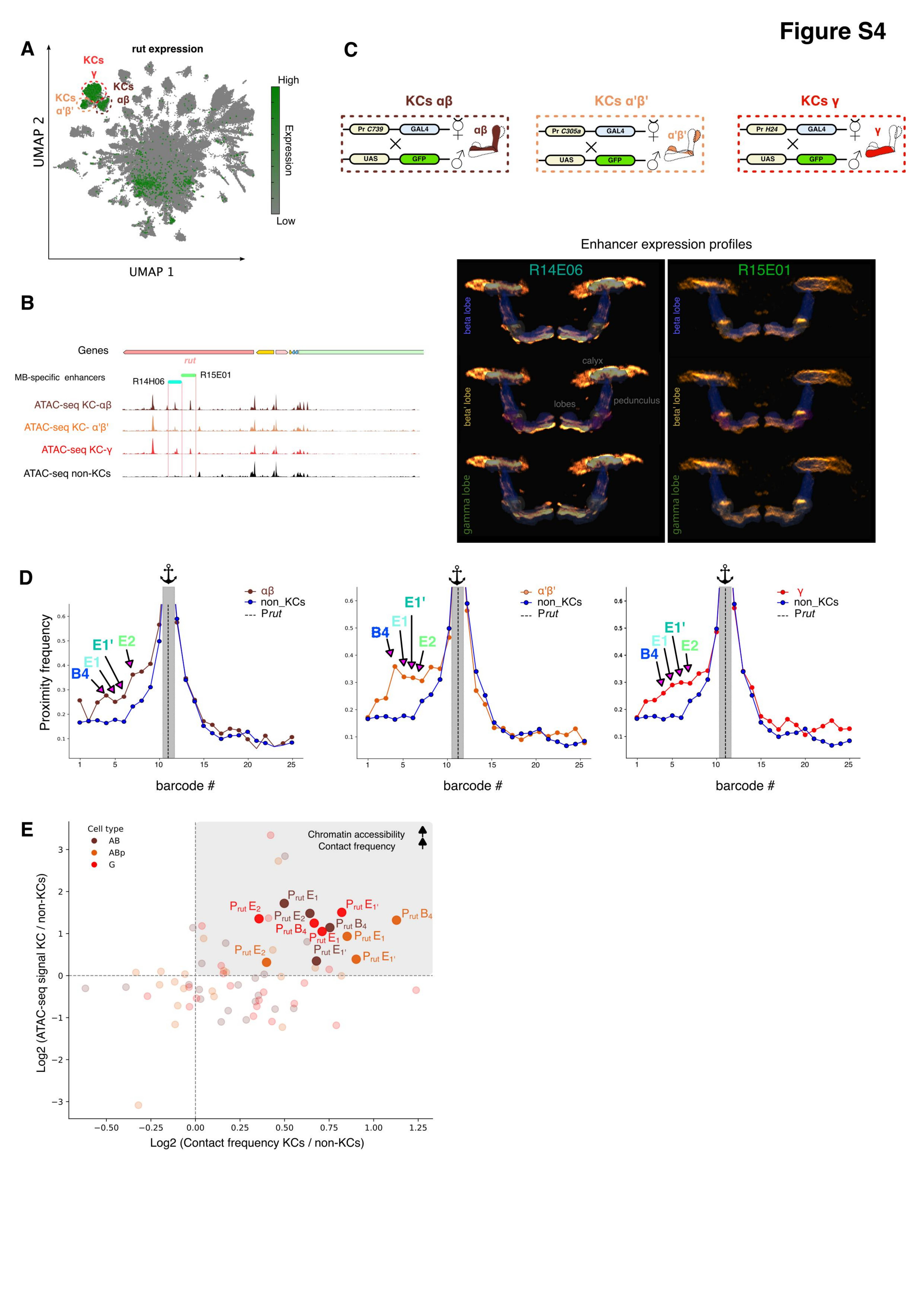}
  \captionsetup{
    width=\textwidth,
    justification=justified,
    singlelinecheck=false
  }
  \vspace{-50pt}
  \caption{%
    \textbf{A} UMAP visualization of adult \textit{Drosophila} brain cell-types based on snRNA-seq data \cite{Davie2018}. Cells expressing textit{rut} are shown in green, cells not expressing \textit{rut} are shown in grey. $\alpha\beta$, $\alpha'\beta'$, and $\gamma$ clusters are highlighted in brown, orange and red. \textbf{B} Left: tracks illustrating  chromatin accessibility (ATAC-seq) profiles in Kenyon cell subtypes and non-KCs. Right: virtual representations of the MB lobes (blue shading) overlapping the expression patterns of the R14E06 and R15E01 enhancers (orange).\textbf{C} Schematic of the UAS-GAL4 system used to drive GFP expression in each KCs subtypes. \textbf{D} Virtual 4M profiles anchored at the \textit{rut} promoter, for each KCs subtypes (brown, orange or red lines) and non-KCs cells (blue line). \textbf{E} Scatter plot showing the differential accessibility of each cell-type compared with non-KCs as a function of their differential contact frequency with the \textit{rut} promoter relative to non-KCs.}
  \label{fig:figS4}
\end{figure}
\clearpage

\end{document}